\documentclass[aps,reprint,superscriptaddress,floatfix,pra]{revtex4-1}

\usepackage{graphicx}
    \graphicspath{{./Images/}}
\usepackage{xcolor}
\usepackage[colorlinks]{hyperref}
\usepackage{enumitem}
    \setlist[itemize]{noitemsep, topsep=0pt}
    \setlist[enumerate]{noitemsep, topsep=0pt}
\usepackage{verbatim}
\usepackage{mathtools,amssymb,amsmath}

\usepackage{bm}
\usepackage{yhmath}
\usepackage[inline]{asymptote}
\usepackage{cancel}
\usepackage{relsize}
\usepackage{array}
\usepackage{bm}

\newcommand{\sete}{\mathbb E}
\DeclareMathOperator{\Var}{Var}

\newcommand{\vt}{\vec{\theta}}
\newcommand{\vdel}{\vec\del}

\newcommand{\add}{\hphantom{===}}
\newcommand{\del}{\Delta}
\newcommand{\ttotal}{t_\text{total}}
\newcommand{\ntotal}{N_\text{total}}
\DeclarePairedDelimiter{\norm}{\Vert}{\Vert}
\let\vec\bm

\newcommand{\ket}[1]{|#1\rangle}
\newcommand{\pdfrac}[2]{\frac{\partial #1}{\partial #2}}

\newcommand{\abs}[1]{\left| #1 \right|}
\begin{document}
\title{Heisenberg-Scaling Measurement Protocol for Analytic Functions with Quantum Sensor Networks}
\author{Kevin Qian}
    \affiliation{Joint Quantum Institute, NIST/University of Maryland, College Park, MD 20742, USA}
    \affiliation{Joint Center for Quantum Information and Computer Science, NIST/University of Maryland, College Park, MD 20742, USA}
    \affiliation{Montgomery Blair High School, Silver Spring, MD 20901, USA}
\author{Zachary Eldredge}
    \affiliation{Joint Quantum Institute, NIST/University of Maryland, College Park, MD 20742, USA}
    \affiliation{Joint Center for Quantum Information and Computer Science, NIST/University of Maryland, College Park, MD 20742, USA}
\author{Wenchao Ge} 
    \affiliation{Institute for Quantum Science and Engineering (IQSE) and Department of Physics \& Astronomy, Texas A\&M University, College Station, Texas 77843, USA}
\author{Guido Pagano}
    \affiliation{Joint Quantum Institute, NIST/University of Maryland, College Park, MD 20742, USA}
    \affiliation{Joint Center for Quantum Information and Computer Science, NIST/University of Maryland, College Park, MD 20742, USA}
\author{Christopher Monroe}
    \affiliation{Joint Quantum Institute, NIST/University of Maryland, College Park, MD 20742, USA}
    \affiliation{Joint Center for Quantum Information and Computer Science, NIST/University of Maryland, College Park, MD 20742, USA}
    \affiliation{IonQ, Inc., College Park, MD 20740}
\author{J.V. Porto}
    \affiliation{Joint Quantum Institute, NIST/University of Maryland, College Park, MD 20742, USA}
\author{Alexey V. Gorshkov}
    \affiliation{Joint Quantum Institute, NIST/University of Maryland, College Park, MD 20742, USA}
    \affiliation{Joint Center for Quantum Information and Computer Science, NIST/University of Maryland, College Park, MD 20742, USA}

\begin{abstract}
We generalize past work on quantum sensor networks to show that, for $d$ input parameters, entanglement can yield a factor $\mathcal O(d)$ improvement in mean squared error when estimating an analytic function of these parameters. We show that the protocol is optimal for qubit sensors, and conjecture an optimal protocol for photons passing through interferometers. Our protocol is also applicable to continuous variable measurements, such as one quadrature of a field operator. We outline a few potential applications, including calibration of laser operations in trapped ion quantum computing.
\end{abstract}

\maketitle
\section{Introduction}

Entanglement is a valuable resource for quantum technology. In metrology, entangled probes are capable of more accurate measurements than unentangled probes \cite{Bollinger1996,Huelga1997,Paris2008,Pezze2009,Toth2012,Zhang2014a}. In addition to using entangled probes to enhance the measurement of a single parameter, using entanglement to estimate many parameters at once, or a function of those parameters, has recently been an area of interest due to  potential applications in tasks such as nanoscale nuclear magnetic resonance imaging \cite{Genoni2013,Humphreys2013,Gao2014,Vidrighin2014,Yue2014,Zhang2014,Kok2017,Proctor2018,Eldredge2018}. 

In this work, we are interested in generalizing the work of Ref.~\cite{Eldredge2018}, which demonstrated a lower bound on the variance of an estimator of a linear combination of $d$ parameters coupled to $d$ qubits. We will generalize this approach to measuring an arbitrary real-valued, analytic function of $d$ parameters and show that entanglement can reduce the variance of such an estimate by a factor of $\mathcal{O} (d)$. Finally, we present a protocol which achieves optimal variance asymptotically in the limit of long measurement time. In addition, when the parameters are coupled to $d$ interferometers or to a combination of interferometers and qubits, we propose an analogous Heisenberg-scaling protocol to improve measurement noise. However, in this case, we lack a proof of optimality. We also can use the protocol presented in Ref.~\cite{Zhuang2018} to couple the parameters to continuous variables detected by homodyne measurements.

We will also examine the application of such a protocol to field interpolation. Suppose sensors are placed at $d$ spatially separated locations, but we wish to know the field at a point with no sensor. We may pick a reasonable ansatz for the field with no more than $d$ parameters, use our $d$ measurements to fix the degrees of freedom of that ansatz, and compute the field at our desired point. Because the field of interest is a function of the field at $d$ other locations, our protocol offers reduced noise over performing the same procedure without using entanglement.  

 \section{Setup}In this work, bold is used to indicate vectors, hats (as in $\hat H$) indicate operators, and variables with a tilde (such as $\tilde f$) are estimators of the corresponding quantity with no tilde (such as $f$). The notation $\sete_Y[X]$ means the expected value of $X$ over all possible $Y$. If we merely write $\sete[X]$, then we average over all parameters required to define $X$ (e.g. if $Y$ depended on $Z$, then $\sete_Z[\sete_Y[X]]$). We define the variance, $\Var_Y[X]$, similarly. 

\begin{figure}
    \includegraphics[width=.45\textwidth]{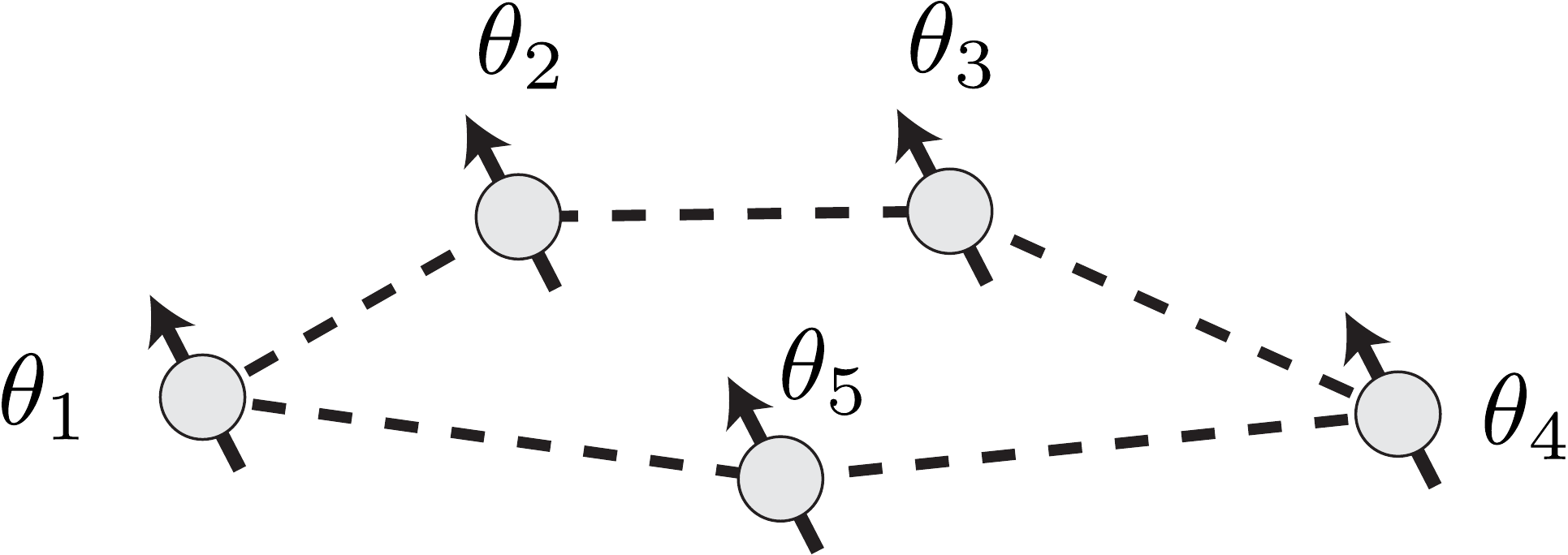}
    \caption{An illustration of a quantum sensor network of spatially separated nodes. At each node, there is an unknown parameter $\theta_i$ coupled to a qubit, which accumulates phase proportional to $\theta_i$.}
    \label{spin schematic}
\end{figure}

We consider a system with $d$ sensor nodes, where node $i$ consists of a single qubit coupled to a real parameter $\theta_i$ (see Fig.~\ref{spin schematic}), and suppose that the state evolves under the Hamiltonian
\begin{equation}
    \hat H = \hat H_c(t) + \frac12 \theta_i \hat \sigma_i^z, \label{hamiltonian}
\end{equation}
where $\hat \sigma_i^{x, y, z}$ are the Pauli operators acting on qubit $i$ and $\hat H_c(t)$ is a time-dependent control Hamiltonian that we choose, which may include coupling to ancilla qubits. Here, and throughout the paper, repeated indices indicate summation. We want to measure an arbitrary real-valued, analytic function $f(\vt)$ of $d$ unknown parameters $\vt = \langle \theta_1, \dots, \theta_d \rangle$ for time $\ttotal$. We would like to determine how well the quantity $f(\vt)$ can be estimated, and find a protocol for doing so. To specify a protocol, we choose an input state, a control Hamiltonian $\hat H_c(t)$, and a final measurement. 

For a general estimator, we use the mean squared error (MSE) $M$ of our estimate $\tilde f$ from the true value $f(\vt)$ as a figure of merit. Explicitly, 
\begin{equation} \label{MSE biased bound} 
    M = \sete[(\tilde f - f(\vt))^2] = \Var \tilde f + (\sete[\tilde f] - f(\vt))^2.
\end{equation}
Thus the MSE accounts for both the variance and the bias of the estimator $\tilde f$. \textcolor{black}{By proving lower-bounds for $M$ and then showing that these bounds are saturable, we will be demonstrating protocols which are optimal in this combination of bias and variance.}

{\color{black}
\section{Lower bound on error}We now identify the minimum possible error of an estimator of $f(\vt)$ which measures for time $\ttotal$. For any estimator $\tilde f$, biased or otherwise, which uses samples from a probabilistic process (such as physical experiments) to estimate the value $f(\vt)$, the MSE is bounded by \cite{Braunstein1994}
\begin{equation} 
    \label{eqn:tmcrb}
    \sete[(\tilde f - f(\vt))^2] \geq \frac{1}{F} \geq \frac{1}{F_Q},
\end{equation}
where $F$ is the Fisher information for the parameter $f$ and $F_Q$ is the quantum Fisher information evaluated over our input state, with $F_Q \geq F$ always \cite{Baumgratz2015}. Bounds on the error of an estimator in terms of the Fisher information are known as Cram\'er-Rao bounds. The Fisher information measures the sensitivity of the sampled probability distribution to changes in the parameters $\vec{\theta}$. While $F$ tells us something about a particular experimental setup, $F_Q$ is maximized over all possible experiments that could be performed on a state.

In order to evaluate the Fisher information for our function of interest $f$, we will use the method presented in Ref.~\cite{Boixo2007} and developed for linear functions in Ref.~\cite{Eldredge2018}. We start by evaluating the generator $\hat{g} = \partial \hat{H}/\partial f$ as defined in Ref.~\cite{Boixo2007}. By first writing the chain rule, we find that
\begin{equation}
	\hat{g} = \pdfrac{\hat{H}}{f} = \pdfrac{\hat{H}}{\theta_i} \pdfrac{\theta_i}{f} = \frac{1}{2} \hat{\sigma}^i_z \pdfrac{\theta_i}{f}.
	\label{eqn:generator}
\end{equation}
Note that $F_Q$ can be upper-bounded by the seminorm of this generator, $F_Q \leq t^2 \| \hat{g} \|_s^2$ \cite{Boixo2007}. (The seminorm of an operator is the difference between its maximum and minimum eigenvalues.) However, to evaluate the seminorm, we will need to evaluate the partial derivative in Eq.~\eqref{eqn:generator}. To do so we must specify a full basis of functions so that the partial derivative can be defined, which requires specifying which variables are held constant during differentiation. We suppose that a set of functions $f_1, f_2, f_3 \dots f_N$ are created, with the $f$ of interest equal to $f_1$, defining an invertible coordinate transformation on a region $\mathbb{R}^N$ around the point $\mathbf{\theta}$. The seminorm is then:
\begin{equation}
	\| \hat{g} \|_s = \sum_{i=1}^N \left| \pdfrac{\theta_i}{f} \right| = \sum_{i =1}^N \abs{J_{i 1}^{-1}}.
\end{equation}
Here, $J_{i j}^{-1}$ is an element of the Jacobian matrix of the inverse transformation to that defined by the $f$ functions. Depending on which functions are chosen, the value of $\| \hat{g} \|_s$ can vary, as can be seen in Ref.~\cite{Eldredge2018} for linear functions. We therefore wish to find the smallest possible $\| \hat{g} \|_s$, which will provide the tightest possible bound on $F_Q$. To do so, we note that $J^{-1}$ and $J$ must obey an inverse relationship, meaning that the following chain of inequalities holds,
\begin{equation}
	1 = J_{1 i} J^{-1}_{i 1} \leq \abs{J_{1 i}} \abs{J^{-1}_{i 1}} \leq \max_j \abs{J_{1 j}} \sum_{i=1}^N \abs{J^{-1}_{i 1}}.
\end{equation}
By using the definition of the Jacobian, we can rewrite this as a lower bound on the value of $\| g \|_s$ in terms of partial derivatives of $f$:
\begin{equation}
	\| \hat{g} \|_s = \sum_{i=1}^N \abs{J_{ i 1}^{-1}} \geq \left( \max_j \left| \pdfrac{f}{\theta_j} \right| \right)^{-1}.
	\label{eqn:genlowerbound}
\end{equation}
All that remains is to note that if we label the $\theta_i$ that yields the maximum first derivative as $\theta_1$, and then choose $f_i = \theta_i$ for $ i > 1$, the lower bound in Eq.~\eqref{eqn:genlowerbound} is met, since $\partial \theta_i/\partial f_1$ must be evaluated holding the other $f_j$ constant. Invoking the resulting bound on the quantum Fisher information, we find that the quantum Cram\'er-Rao bound becomes
\begin{equation}
	M = \sete[(\tilde f - f(\vt))^2] \geq \frac{1}{F_Q} \geq \max_j \frac{ \abs{\pdfrac{f}{\theta_j}}^2}{t^2}.
	\label{time lower bound}
\end{equation}

Although the quantum Cram\'er-Rao bound derived in Eq.~\eqref{time lower bound} cannot always be saturated, it can when the generators $\partial \hat H/\partial\theta_i$ commute, as in Eq.~(\ref{hamiltonian}) \cite{Baumgratz2015}. We will show later that the inequality in Eq.~\eqref{time lower bound} can be saturated at asymptotic time $\ttotal$. 

From this point forward, to simplify later calculation, we define $f_i(\vt) = \frac{\partial f(\vt)}{\partial \theta_i}$. This definition also generalizes to multiple partial derivatives (i.e. $f_{ij} = \frac{\partial}{\partial \theta_j} \frac{\partial f}{\partial \theta_i}$). 

Before moving on to the optimal protocol, we will consider a protocol which does not use entanglement and does not saturate Eq.~(\ref{time lower bound}) as a useful contrast to an entangled strategy. Suppose we estimate each parameter individually, without bias. Then the MSE $\sete[(f(\tilde\vt) - f(\vt))^2]$ can be written as
\begin{equation}
    M_\mathrm{unentangled} = f_i(\vt)^2 \Var\tilde\theta_i. \label{naive general FoM}
\end{equation}
Here we assume the measurement of each single parameter can be made in time $t$ with variance $\Var\tilde\theta_i = \frac 1{t^2}$, \textcolor{black}{the Heisenberg limit for single particles and therefore the best possible measurement for a non-entangled protocol \cite{Bollinger1996}. Estimation protocols that allow one to reach a variance proportional to $1/t^2$ without entanglement are outlined in detail in Ref.~\cite{Kimmel2015}; an experimental realization of single phase estimation without entanglement was performed in Ref.~\cite{Rudinger2017}. While in realistic settings a Heisenberg-limited measurement on one particle may be challenging and include some constant overhead above $1/t^2$, this assumption allows us to identify the improvement possible by using entanglement.} Our entanglement-free figure of merit is
\begin{equation}
    M_\mathrm{unentangled} = \frac{\norm*{\nabla f(\vt)}^2}{\ttotal^2}, \label{naive time FoM}
\end{equation}
where the $\| \cdot \|$ in Eq.~\eqref{naive time FoM} denotes the Euclidean norm. More generally, we use $\norm*{\vec v}_p$ to denote the $p$-norm of vector $\vec v$. Since Eq.~\eqref{naive time FoM} only saturates Eq.~\eqref{time lower bound} in trivial cases where $\nabla f(\vt)$ is zero in all but one component, the unentangled protocol described is not optimal. 
}

 \section{Two-step Protocol}We now present a protocol which asymptotically saturates Eq.~\eqref{time lower bound}. Our protocol consists of two steps. First, we make an unbiased estimate $\tilde\vt$ of $\vt$ for time $t_1$. Second, given our estimates $\tilde\vt$, we make an unbiased measurement $\tilde q$ of the quantity $q = \nabla f(\tilde\vt) \cdot (\vt - \tilde\vt)$ using the linear combination protocol in Ref.~\cite{Eldredge2018}, which takes time $t_2$. Our final estimate is $\tilde f = f(\tilde\vt) + \tilde q$. 

It can be shown that our protocol is optimal (in terms of scaling with the total time $t_1$ + $t_2$) provided that the individual estimations of the parameters satisfy $\sete[(\tilde{\theta}_i - \theta_i )^4] = \mathcal O(t_1^{-4})$ and that $t_1$ and $t_2$ are chosen properly. To simplify our computations, we will make the more concrete assumption that our initial estimates $\tilde\vt$ are each normally distributed as $\mathcal N(\theta_i, \Var\tilde\theta_i)$. Then as computed in the Appendix, the figure of merit for this protocol is 
\begin{align}
    \!\!\!\!M 
    &= \sete[(f(\tilde\vt) + \tilde q - f(\vt))^2] \\
   &= \sete[\Var_{\tilde q} \tilde q] 
    + \frac{2f_{ij}(\vt) + f_{ii}(\vt)f_{jj}(\vt)}4 \Var\tilde\theta_i \Var\tilde\theta_j.
    \label{general FoM}
\end{align}
In Eq.~\eqref{general FoM}, the first term is the error resulting from the second phase of the protocol, estimating the linear combination. The second term is a residual error remaining from the first phase of the protocol after it is corrected by the linear combination measurement.

For our particular Hamiltonian $\hat H = \frac12 \theta_i \hat \sigma_i^z$, as per Ref.~\cite{Eldredge2018}, we know that the minimum variance of an unbiased estimator of some linear combination $\vec\alpha \cdot \vt$ given time $t$ is 
\begin{equation}
    \Var \widetilde{\vec\alpha \cdot \vt} \ge \frac{\max_i \alpha_i^2}{t^2},
\end{equation}
which can be achieved with the entangled GHZ state $\ket{\psi_\mathrm{spin}} = \frac{1}{\sqrt{2}} (\ket{0}^{\otimes d} + \ket{1}^{\otimes d})$. We can apply this linear combination protocol to the second phase of our protocol by setting $\vec \alpha = \nabla f(\tilde\vt)$. For the individual estimators of the first phase, we use the fact that an individual estimation can be made in time $t$ with variance $1/t^2$ \cite{Bollinger1996}. Using these results, we simplify Eq.~\eqref{general FoM}: 
\begin{align}
    M &= \sete\left[\frac{\max_i f_i(\tilde\vt)^2}{t_2^2}\right] + \frac{\frac{2f_{ij}(\vt) + f_{ii}(\vt) f_{jj}(\vt)}4}{t_1^4} \\
    &= \frac{\sete[\max_i f_i(\tilde\vt)^2]}{t_2^2} + \frac{g_1(\vt)}{t_1^4}, \label{incomplete time MSE}
\end{align}
where we have absorbed the second derivatives into $g_1(\vt)$, which does not depend on time. Without loss of generality, we designate $f_1(\tilde\vt)$ as the largest $f_i (\tilde{\vec{\theta}})$. We then expand $\sete[f_1(\tilde\vt)^2]$ as
\begin{equation}
    f_1(\vt)^2 
        + \frac{f_1(\vt)f_{1ii}(\vt)}{t_1^2} 
        + \frac{f_{1i}(\vt)^2}{t_1^2} 
        + \mathcal O((\tilde\vt - \vt)^3). \label{E of max squared}
\end{equation}
We may substitute Eq.~(\ref{E of max squared}) into Eq.~(\ref{incomplete time MSE}) to obtain
\begin{equation}
    M = \frac{g_2(\vt)}{t_2^2} + \frac{g_3(\vt)}{t_1^2t_2^2} + \frac{g_1(\vt)}{t_1^4} + \mathcal O((\tilde\vt - \vt)^3), \label{time MSE}
\end{equation}
where $g_2(\vt) = f_1(\vt)^2$ and $g_3(\vt)$ have been introduced to absorb more time-independent factors. 

\subsection{Optimal time allocation}
To complete the protocol, we must specify how the total time $\ttotal$ is to be allocated between $t_1$ and $t_2$. We want to choose the $t_1, t_2$, under the constraint that $t_1 + t_2 = \ttotal$, which minimize the MSE 
\begin{equation}
    M = \frac{g_2(\vt)}{t_2^2} + \frac{g_3(\vt)}{t_1^2t_2^2} + \frac{g_1(\vt)}{t_1^4}. 
\end{equation}
Notice that the $g_1, g_2, g_3$ functions are only dependent on $\vt$ and not $t_1$, so we may set the derivative of $M$ with respect to $t_1$ equal to $0$ and obtain
\begin{equation} 
    \frac{2g_2(\vt)}{t_2^3} + \frac{2g_3(\vt)}{t_2^3t_1^2} 
    = \frac{2g_3(\vt)}{t_2^2t_1^3} +  \frac{4g_1(\vt)}{t_1^5}. 
\end{equation}
Let $r = t_1/t_2$. Then we may rearrange to obtain
\begin{equation} 
    g_2(\vt)t_1^2 
    = \frac{g_3(\vt)}r + \frac{2g_1(\vt)}{r^3} - g_3(\vt). 
\end{equation}
Since $t_1 \gg 1$, then $r \ll 1$, so the $r^{-3}$ term dominates the RHS. Thus,  $g_2(\vt)t_1^2 \approx \frac{2g_1(\vt)}{r^3}$, which implies
\begin{equation}
    t_1 
    \approx \left(\frac{2g_1(\vt)}{g_2(\vt)}\right)^{1/5} t_2^{3/5} \approx \left(\frac{2g_1(\vt)}{g_2(\vt)}\right)^{1/5}\ttotal^{3/5}. \label{t1 formula}
\end{equation} 
Therefore the best possible allocation satisfies
\begin{equation}
    t_1 = g(\vt) \ttotal^{3/5}, \label{eq:optimal allocation}
\end{equation}
where $g$ is a function which depends only on $f$ and $\vt$. In particular, $t_1 = \mathcal O(\ttotal^{3/5})$, so the fraction of time spent on $t_1$ vanishes as $\ttotal \to \infty$. Almost all of the time is spent on $t_2$, the linear combination step of the two-step protocol. It can readily be shown that Eq.~\eqref{time MSE} is asymptotically dominated by the first term when this time allocation is chosen, which (since $t_2 \to \ttotal$) is equal to the right-hand-side of the bound in  Eq.~\eqref{time lower bound}. In other words, this distribution of time asymptotically achieves the optimal MSE.

The two-step protocol exhibits Heisenberg scaling as defined for distributed sensing \cite{Eldredge2018,Ge2018,Proctor2018}.
Comparing Eq.~(\ref{naive time FoM}) to Eq.~(\ref{time lower bound}) shows an improvement of $\mathcal O(d)$, maximized when all components of $\nabla f(\vt)$ are approximately equal. Intuitively, the advantage is maximal when all parameters contribute, but minimal (i.e.\ no advantage) when only one parameter affects the function value. Similar behavior was noted in the linear combination case \cite{Eldredge2018}. 

Note that when actually implementing the protocol, the optimal $t_1$ is unknown since the function $g$ that determines it depends on the true parameters $\vt$. However, we do not need to use the optimal $t_1$ to saturate the bound in Eq.~(\ref{time lower bound}). If $t_1$ is a function $c\ttotal^p$ of the total time where $\frac 12 < p < 1$ and some constant $c$, then the protocol will saturate Eq.~(\ref{general FoM}). Suppose that $t_1 = c\ttotal^p$ for some $\frac 12 < p < 1$ and some constant $c$. Since $p < 1$, we see that $\lim_{t \to\infty} \frac{t_2}\ttotal = 1$. Therefore, we may substitute our $t_1$ into the MSE formula in Eq.~(\ref{time MSE}) and simplify:
\begin{equation}
    \lim_{\ttotal \to \infty} M = \lim_{\ttotal \to \infty} \frac{g_2(\vt)}{\ttotal^2} + \frac{g_3(\vt)}{c^2\ttotal^{2+2p}} + \frac{g_1(\vt)}{c^4\ttotal^{4p}}. \label{time MSE for p}
\end{equation}
Since $p > \frac 12$, the $\ttotal^2$ term is dominant. Thus, as we defined $g_2 := f_1(\vt)^2 = \max_i f_i(\vt)^2$ under the assumption that $f_1(\vt)^2$ was maximal, our asymptotic error is
\begin{equation}
    M = \frac{\max_i f_i(\vt)^2}{\ttotal^2}, \label{rewrite time FoM}
\end{equation}
which saturates the bound of Eq.~\eqref{time lower bound}. Although selecting a non-optimal time allocation does result in a higher MSE, the additional error is  $\mathcal{O}\left(\ttotal^{-4}\right)$, which is insignificant asymptotically. The two-step protocol will therefore be asymptotically optimal for a wide range of time allocations. 

 \section{Function Measurement in Other Physical Settings}
\begin{figure}[h]
    \includegraphics[width=.45\textwidth]{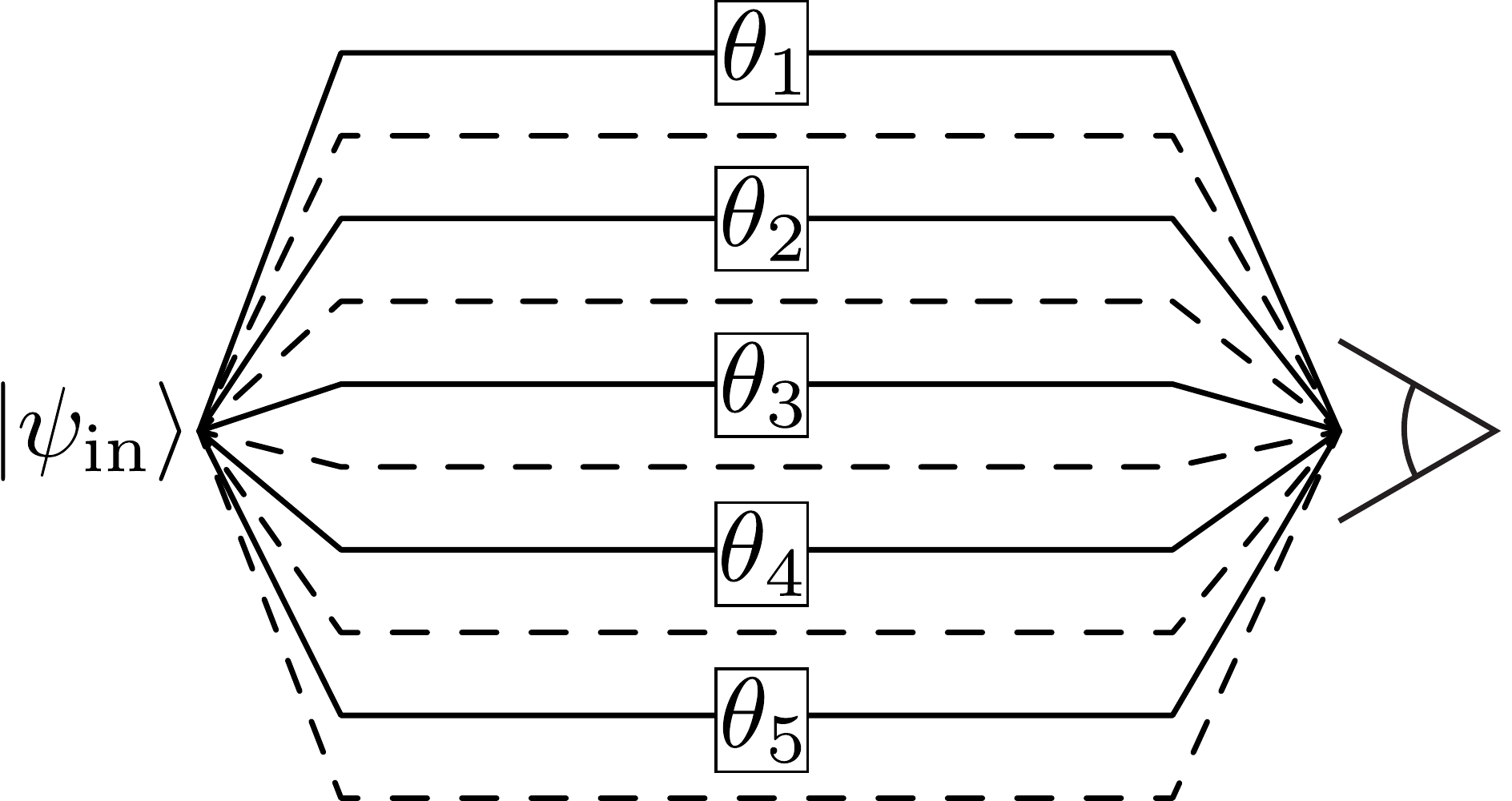}
    \caption{An example illustration of a quantum sensor network composed of separate interferometers. In each, one arm accumulates an unknown phase $\theta_i$ and the other arm is a reference port with no phase. }
    \label{photon schematic}
\end{figure}

We now consider a different physical setting for function estimation. Rather than $d$ qubits which accumulate phase for some time $t$, we instead pass $n$ photons through $d$ Mach-Zehnder interferometers and accumulate some fixed phase $\theta_i$ encoded into each interferometer (see Fig.~\ref{photon schematic}). For single parameters, the use of entangled states to reduce noise in this setting has been explored in Refs.~\cite{Holland1993,Kim1998,UshaDevi2009,Demkowicz-Dobrzanski2015,Dinani2016} with multiparameter cases explored in Refs.~\cite{Ge2018,Proctor2018}. In this setting, the relevant limitation is the total number of photons used in the measurement, rather than time. This constraint is particularly relevant when analyzing a biological or chemical sample which is sensitive to light, making it desirable to reduce noise with as few photons as possible. Similar biologically motivated situations are presented in Refs.~\cite{Kee2004,Alem2015,Jensen2016}.
    
For photons, a two-step protocol with similar structure to the protocol for qubits yields reduced noise compared to any estimate of $f$ derived entirely from local measurements. Suppose we allot $N_1$ photons for the first step (individual measurement) and $N_2$ photons for the second step (linear combination), for a total of $\ntotal = N_1 + N_2$ photons. We again begin from the general result of Eq.~(\ref{general FoM}). However, the use of photons which can be apportioned between modes introduces new structure to the problem. We need to partition the $N_1$ photons into $N_1 = n_1 + \dots + n_d$, putting $n_i$ photons into the $i$-th interferometer, as some parameters may affect our final result more than others. Thus,  in the second term of Eq.~(\ref{general FoM}), we replace $\Var\tilde\theta_i$ with $\frac1{n_i^2}$ instead of  $\frac1{t_1^2}$ \cite{Holland1993}. 

The optimal variance when measuring the linear combination $\vec\alpha \cdot \vt$ using $N$ total photons is unknown. However, Ref.~\cite{Proctor2018} conjectures the optimal variance to be
\begin{equation} 
    \Var \widetilde{\vec\alpha \cdot \vt} \ge \frac{\norm*{\vec\alpha}_1^2}{N^2}. \label{photon lower bound}
\end{equation}
Furthermore, Ref.~\cite{Proctor2018} provides a protocol achieving the bound in Eq.~(\ref{photon lower bound}) using a proportionally weighted GHZ state: $\ket{\psi_{ \mathrm{photon} }} = \frac{1}{\sqrt{2}} ( \ket{n_1, 0, n_2, 0 \dots} + \ket{0, n_1, 0, n_2, \dots})$, where $n_i = \ntotal \frac{\alpha_i}{\sum \alpha_j}$ and where, in reference to Fig.\ \ref{photon schematic}, the modes are listed from top to bottom. Note that this will only work for $\vec\alpha$ proportional to some rational vector as photons are discrete. Since Eq.~(\ref{photon lower bound}) is saturable, we may simplify the first term of Eq.~(\ref{general FoM}) to obtain 
\begin{align}
    M &= \frac{\sete \left[\norm*{\nabla f(\tilde\vt)}_1^2 \right]}{N_2^2} + \frac{2f_{ij}(\vt)^2 + \frac{f_{ii}(\vt) f_{jj}(\vt)}4}{n_i^2n_j^2}.  \label{photon MSE}
\end{align}
For fixed $f$ and $\theta$, the $\frac1{n_in_j}$ terms in Eq.~(\ref{photon MSE}) are minimized for the same ratio of $n_1: n_2 : \dots : n_d$ regardless of the value of the total number of photons used, $N_1$. Each term is proportional to $N_1^{-4}$ multiplied by some function of $f, \vt$, and $d$. Therefore, the structure of Eq.~\eqref{photon MSE} becomes identical to the structure of Eq.~\eqref{time MSE}, with $N_1$ and $N_2$ replacing $t_1$ and $t_2$. As a result, the optimal allocation of photons between $N_1$ and $N_2$ will yield $N_1 = \mathcal O(\ntotal^{3/5})$ and $N_2 = \mathcal O(\ntotal)$, meaning that the $N_2^{-2}$ term in Eq.~(\ref{photon MSE}) is dominant asymptotically. Therefore, for photons, we may asymptotically achieve
\begin{equation}
    M = \frac{\norm*{\nabla f(\tilde\vt)}_1^2}{\ntotal^2} + \mathcal O \left( \frac1{\ntotal^{12/5}} \right). \label{2 step photon FoM}
\end{equation}
This strategy is optimal if the linear combination estimation strategy presented in Ref.~\cite{Proctor2018} is optimal, as conjectured in that work. \textcolor{black}{We stress that our optimality result remains true for spins evolving under Eq.~\eqref{hamiltonian} and it is only for photons that our protocol is only conjectured to be optimal.}

Eq. (\ref{2 step photon FoM}) also exhibits Heisenberg scaling. Suppose we were to measure each parameter individually and then calculate the function. When measuring the parameters individually, we obtain the same error formula as Eq.~(\ref{naive general FoM}), except now we set $\Var\tilde\theta_i = \frac1{n_i^2}$ to get
\begin{equation}
    M_\mathrm{unentangled} = \frac{f_i(\vt)^2}{n_i^2}.
\end{equation}
The optimal distribution requires an $n_i$ proportional to the weight $f_i(\vt)^{2/3}$, yielding an entanglement-free error of 
\begin{equation}
    M_\mathrm{unentangled} = \frac{\norm*{\nabla f(\vt)}_{2/3}^2}{\ntotal^2}. \label{naive photon FoM} 
\end{equation}
As with qubits, by comparing Eq.~(\ref{2 step photon FoM}) with Eq.~(\ref{naive photon FoM}) in the case where all of the $f_i(\vt)$ are approximately equal, we find that the photonic two-step protocol yields a $\mathcal O(d)$ improvement in error over measuring each parameter individually. This improvement when all quantities are equally important can also be seen in Ref.~\cite{Ge2018} for the special case of $f$ being a linear combination. As in the qubit case, the improvement in error is lessened when $\nabla f(\vt)$ is not approximately equal in all components. 

In fact, this method can be extended still more generally. Rather than cases where the signal is imprinted on photons by a phase shift, we can consider the protocol developed in Ref.~\cite{Zhuang2018}, which is capable of entanglement-enhanced distributed sensing of continuous variables by using homodyne measurements. Besides measuring parameters in different physical settings, we may also measure functions of variables coupled to spins, phase-shifts of photons, continuous variables, and any combination of these. In such a hybrid scenario, we can still make use of the two-step protocol. The first step, obtaining initial estimates for the individual parameters, proceeds equivalently, since the measurements of the spins and of the photons can be viewed as occurring in parallel. For the linear combination case, we can assume that the optimal spin and photon input states can be entangled as follows:
\begin{align} \label{eq:spin-photon}
\ket{\psi_{ \mathrm{spin-photon} }} = \frac{1}{\sqrt{2}} \big( &\ket{n_1, 0, n_2, 0 \dots} \otimes \ket{1,1,1,\dots} \\ \nonumber + &\ket{0, n_1, 0, n_2, \dots} \otimes \ket{0,0,0,\dots} \big). \end{align}
Here, $n_i = \ntotal \frac{\alpha_i}{\sum \alpha_j}$, where the sum runs over only the $j$ corresponding to photonic modes, denotes the number of photons which pass through the arms of the $i$-th interferometer. The state in Eq.\ (\ref{eq:spin-photon}) is designed in such a way that the two branches of the overall wavefunction accumulate relative to each other a phase equal to the total linear combination we are interested in. In order to extract this final phase, the state can be unitarily mapped onto a qubit, which contains all of the accumulated phase and is then measured.

One caveat is that the linear combination protocol will accumulate phase proportional to time for the qubits and phase proportional to the number of photons for interferometers. For instance, if $\theta_1$ is coupled to a qubit (and therefore has units of frequency) and $\theta_2$ is coupled to an interferometer (and is therefore unitless), then the two branches of our state accumulate a relative phase $\theta_1t + \theta_2n$. Therefore, one may have to adjust $t$ or $n$ in order to get the desired linear combination. 

\section{Applications}\textcolor{black}{Our protocol is capable of estimating any analytic function of the inputs, allowing for a large variety of potential applications. Essentially, any time multiple sensors are processed into a single signal, our protocol provides enhanced sensitivity using entanglement.} In fact, there is no requirement that different $\theta_i$ have the same physical origin. For instance, a $\theta_1$ representing an electric field and $\theta_2$ measuring a magnetic field could be used to measure the Poynting vector. 

One potential application of function measurements is the interpolation of non-linear functions. Suppose that an ansatz with $d$ tunable parameters is made for the strength of the field in a region. With readings from $\geq d$ different points, one could determine the parameters of the ansatz and therefore determine the value of the field at other points. Estimations of these ansatz parameters, which are functions of the measured fields, may potentially be improved using entangled states depending on the figure of merit \cite{Chiribella2005, Baumgratz2015}. Note that this procedure can be carried out even if it is difficult to invert the ansatz in terms of the $d$ measurements. Suppose that $\vt = f(\vec c, \vec x)$ and that $\vec c = f^{-1}(\vt, \vec x)$ exists, but has no closed-form solution which can be easily evaluated. First, we make measurements $\hat\vt$. To create an initial estimate of the values $\vec c$, we use a numerical root-finder to find estimates $\tilde{\vec c}$. We can now implement the second step of our protocol by finding the first derivatives $\partial c_i/\partial \theta_j$ using the matrix identity $\frac{\partial \vt}{\partial \vec c} \cdot \frac{\partial \vec c}{\partial \vt} = I$. Since $f$ is known, $\partial \vt/\partial \vec c$ can be inverted to yield the $\partial \vec c/\partial \vt$ needed to estimate $\hat{\vec q} = \partial \vec c/\partial \vt |_{\vt = \hat\vt} \cdot (\vt - \hat\vt)$. Our final estimate is $\hat{\vec c} + \hat{\vec q}$, which was obtained without having to compute $f^{-1}$ in general.

Interpolation in this manner can proceed by two different schemes. We can either attempt to measure the ansatz parameters themselves, which allows computation of the field at all other points, or we can skip the final computation step by writing the field at a point of interest as a function of all the points that can be measured. This final function can then be directly measured using an entangled protocol, which will be more accurate. However, the first approach has the advantage that knowing the ansatz parameters allows estimation of all points in the space in question.

One particular interpolation of interest arises in ion trap quantum computing. In trapped ion chains, qubits are manipulated using Gaussian laser beams, and two primary sources of error are intensity and beam pointing fluctuations \cite{Cirac1995,Haffner2008,Brown2011}. Our protocol offers better ways to characterize this noise. In order to detect the field error at a qubit's position without disturbing the qubit, we can perform interpolation by measuring the field's effect on other ions, possibly of a different atomic species, positioned nearby. Given the ansatz of the Gaussian beam profile, we are able to calculate the field at the qubit of interest and perhaps correct the error. \textcolor{black}{As entanglement of ions is already a key functionality for trapped ion quantum computers, our proposal is immediately applicable in that domain.} 

\section{Outlook}We have presented a Heisenberg-scaling measurement protocol using quantum sensor networks for measuring any multivariate, real-valued, analytic function, and this protocol is consistent with the Heisenberg limit when measuring functions with comparably-sized gradients in each component. \textcolor{black}{Recent advances in the distribution of entanglement, for instance, in satellites distributing entangled photons more than 1000km \cite{Yin2017}, strengthen the viability of this scheme over large distances in the near-term. Potential sensing platforms include trapped ions and nitrogen-vacancy defects in diamond, which can also be entangled \cite{Leibfried2004,Bernien2013,Dolde2014, Hensen2015} and are proven platforms for magnetometry and thermometry \cite{Taylor2008,Tzeng2015}.} Future work may include proving the optimality of the two-step protocol when constrained by the number of photons, which would require extending the results of Ref.~\cite{Proctor2018}, \textcolor{black}{as well as further experimental research into quantum networking to explore how entanglement can be reliably distributed for metrological purposes.}

We specifically identified field interpolation as a promising application of our work, but we stress that our protocol can assist in the measurement of any analytic function. More work remains to determine when it is optimal to measure the coefficients of interpolation and when it is optimal to directly measure the final function. We are also interested in fleshing out possible intersections between quantum function estimation and machine learning. Supervised machine learning is a type of interpolation: estimating functional outputs for unknown inputs by extracting information from known input-output pairs \cite{Norvig2010}. It is possible our protocol could be used to improve the accuracy of training a machine learning model if the necessary quantity for training was a function of physical measurements. Additionally, the final output of many machine learning algorithms, such as neural networks, is a non-linear but infinitely differentiable function of the inputs \cite{Schmidhuber2015}. Our work could aid in computing this complicated function for new input when making predictions. 

\begin{acknowledgements}
We would like to thank M.~Foss-Feig, S.~Rolston, J.~Gross, and S.~Kimmel for helpful discussions. This work was supported by ARL CDQI, AFOSR, ARO MURI, ARO, NSF PFC at JQI, NSF PFCQC program, DoE ASCR Quantum Testbed Pathfinder program (Award No. DE-SC0019040), and the DoE BES QIS program (Award No. DE-SC0019449). Z.E. was supported in part by the ARCS Foundation.

\end{acknowledgements}

\bibliography{library}{}

\onecolumngrid
\newpage
\appendix
\section{Figure of Merit for Two-Step Protocol}
\label{calculation of general FoM}
In this section, we derive Eq.~\eqref{general FoM} in the main text. Specifically, we derive the figure of merit for the two-step protocol in terms of the measurement accuracy of the independent parameters and the measurement accuracy of the linear combination, yielding a general formula which applies to any physical realization. 

For the sake of concision, let $\vdel = \tilde\vt - \vt$ which satisfies $\sete[\vdel] = \vec0$. Furthermore, let $T_k$ be $k!$ times the $k$-th term of the Taylor expansion of $f$ (so $T_1 = f_i(\vt)\del_i$, $T_2 = f_{ij}(\vt)\del_i\del_j$, $T_3 = f_{ijk}(\vt) \del_i\del_j\del_k$, etc.). Thus, the Taylor expansion of $f(\tilde\vt)$ would be 
\begin{equation} 
    f(\tilde\vt) 
    = f(\vt) + T_1 + \frac{T_2}2 + \frac{T_3}6 + \dots .
\end{equation}
We compute our figure of merit:
\begin{align}
    M
    &= \sete[(f(\tilde\vt) + \tilde q - f(\vt))^2] \\
    &= \underbrace{\sete[(f(\tilde\vt) - f(\vt))^2]}_{\text{term }1}
    + \underbrace{\sete[\tilde q^2]}_{\text{term }2}
    + 2\underbrace{\sete[f(\tilde\vt)\tilde q]}_{\text{term }3}
    - 2 f(\vt)\sete[\tilde q]  \label{naive protocol} \\ 
    \begin{split} 
    &= \left( \underbrace{
        \sete[T_1^2] 
        + \sete[T_1T_2] 
        + \frac13\sete[T_1T_3] 
        + \frac14\sete[T_2^2] 
        + \mathcal O(\vdel^5)
    }_{\text{term }1} \right) 
    + \left( \underbrace{
        \sete[\Var_{\tilde q} \tilde q] 
        + \sete[q^2]
    }_{\text{term }2} \right) \\ 
    &\add + 2 \left( \underbrace{
        f(\vt) \sete[q] 
        + \sete[T_1q] 
        + \frac12 \sete[T_2q] 
        + \frac16 \sete[T_3q] 
        + \mathcal O(\vdel^5)
    }_{\text{term }3} \right) 
    - 2f(\vt) \sete[q] 
    \end{split} \\
    &= 
    \sete[\Var_{\tilde q} \tilde q]
    + \sete[(q + T_1)^2] 
    + \sete[(q + T_1)T_2] 
    + \frac13\sete[(q + T_1)T_3] 
    + \frac14\sete[T_2^2] 
    + \mathcal O(\vdel^5). \label{intermediate bash}
\end{align}
The actual computation of the labeled terms is rather involved and space consuming, so it is presented in Sec.\ \ref{appendix: labeled terms}). Notice that we may simplify
\begin{align} 
    q + T_1 
    &= \del_i(f_i(\vt) - f_i(\tilde\vt)) \\
    &= -\del_i(f_{ij}(\vt) \del_j + \mathcal O(\vdel^2)) \\
    &= -T_2 + \mathcal O(\vdel^3), 
\end{align}
so Eq.~(\ref{intermediate bash}) evaluates to
\begin{align}
    M
    &= \sete[\Var_{\tilde q} \tilde q] 
    + \sete[T_2^2] 
    - \sete[T_2^2] 
    - \frac 13\sete[T_2T_3] 
    + \frac14\sete[T_2^2]
    + \mathcal O(\vdel^5) \\
    &= \sete[\Var_{\tilde q} \tilde q] 
        + \frac14 \sete[T_2^2]
        + \mathcal O(\vdel^5) 
\end{align}
since $\sete[T_2T_3]$ is $\mathcal O(\vdel^5)$. Now, this simplifies further as
\begin{align} 
    M 
    &= \sete[\Var_{\tilde q} \tilde q] + \frac14\sete[T_2^2] \label{begin time simplify}\\
    &= \sete[\Var_{\tilde q} \tilde q] + \frac14 \sete[(f_{ij}(\vt) \del_i\del_j)^2] \\
    &= \sete[\Var_{\tilde q} \tilde q] + \frac14 \sete \left[ 
        4\sum_{i < j} f_{ij}(\vt)^2 \del_i^2 \del_j^2 
        + 2\sum_{i < j} f_{ii}(\vt) f_{jj}(\vt) \del_i^2 \del_j^2 
        + \sum_i f_{ii}(\vt)^2 \del_i^4 
    \right] 
\end{align}
since all terms with some $\del_i$ to a single power will factor out as $\sete[\del_i] = 0$. We will assume that $\del_i \sim \mathcal N(0, \frac1{t_1^2})$ is normally distributed. This is not strictly necessary as long as the distribution of errors satisfies $\sete[\del_i^4] \leq \mathcal{O}(t_1^{-4})$, a condition that is satisfied by phase estimation procedures like those in Ref.~\cite{Kimmel2015}. However, assuming normality allows the calculation to proceed easily, as we will be able to  simplify $\sete[\del_i^4] = 3\Var\tilde\theta_i^2$. Thus, we arrive at
\begin{align}
    M &= \sete[\Var_{\tilde q} \tilde q] + \frac14\left( 4\sum_{i < j} f_{ij}(\vt)^2 \Var \tilde\theta_i \Var \tilde\theta_j + 2 \sum_{i < j} f_{ii}(\vt) f_{jj}(\vt) \Var\tilde\theta_i \Var\tilde\theta_j + \sum_i 3f_{ii}(\vt)^2\Var\tilde\theta_i^2\right) \\
    &= \sete[\Var_{\tilde q} \tilde q] + \sum_{i, j} \frac{2f_{ij}(\vt) + f_{ii}(\vt)f_{jj}(\vt)}4 \Var\tilde\theta_i \Var\tilde\theta_j. 
\end{align}
\subsection{Simplification of labeled terms}
\label{appendix: labeled terms}

In this subsection, we present the simplification of the labeled terms from Eqs.~(\ref{naive protocol}-\ref{intermediate bash}) in full detail. 

Term 2  is simplified by using the definition of $\Var_{\tilde q} \tilde q$. One needs to be careful as there are two layers of expected values - one for the values of $\tilde\vt$ and one for the estimator $\tilde q$: 
\begin{align} 
    \underbrace{\sete[\tilde q^2]}_{\text{term }2} &= \sete_{\tilde\vt}[\sete_{\tilde q}[\tilde q^2]] \\
    &= \sete_{\tilde\vt}[\Var_{\tilde q} \tilde q + \sete_{\tilde q}[\tilde q]^2] \\
    &= \sete_{\tilde\vt}[\Var_{\tilde q} \tilde q + q^2] \\
    &= \sete[\Var_{\tilde q} \tilde q] + \sete[q^2]. 
\end{align}

Terms 1 and 3 are simplified by expanding the Taylor series for $f(\tilde{\theta})$ up to $\vdel^4$ terms; note that $q = \mathcal O(\vdel)$, so we only need to expand the Taylor series up to $\mathcal O(\vdel^3)$ terms: 
\begin{align}
    \underbrace{\sete\left[\left(f(\tilde\vt) - f(\vt) \right)^2 \right]}_{\text{term }1}
    &= \sete[f(\tilde\vt)^2] 
    - 2f(\vt) \sete[f(\tilde\vt)]
    + f(\vt)^2 \\
    \begin{split}
    &= 
        f(\vt)^2 
        + \sete[T_1^2] 
        + f(\vt)\sete[T_2] 
        + \sete[T_1T_2]
        + \frac13 f(\vt) \sete[T_3]
    \\
    &\add 
        + \frac 1{12}f(\vt)\sete[T_4] 
        + \frac13 \sete[T_1T_3] 
        + \frac14 \sete[T_2^2] 
        + \mathcal O(\vdel^5)
    \\
    &\add - 2f(\vt) \left( 
        f(\vt) 
        + \frac12 \sete[T_2] 
        + \frac16\sete[T_3] 
        + \frac 1{24}\sete[T_4] 
        + \mathcal O(\vdel^5)
    \right) 
    + f(\vt)^2
    \end{split} \\
    &= \sete[T_1^2] 
    + \sete[T_1T_2] 
    + \frac13\sete[T_1T_3] 
    + \frac14\sete[T_2^2] 
    + \mathcal O(\vdel^5). 
\end{align}
\begin{align}
    \underbrace{\sete[f(\tilde\vt)\tilde q]}_{\text{term }3}
    &= \sete_{\tilde\vt}[\sete_{\tilde q}[f(\tilde\vt) \tilde q]] \\
    &= \sete_{\tilde\vt}[f(\tilde\vt) q] \\ 
    &= \sete\left[\left(f(\vt) + T_1 + \frac {T_2}2 + \frac{T_3}6 + \mathcal O(\vdel^4) \right) q\right] \\
    &= f(\vt) \sete[q] + \sete[T_1q] + \frac{\sete[T_2q]}2 + \frac{\sete[T_3q]}6 + \mathcal O(\vdel^5). 
\end{align}

\end{document}